*Laboratoire de l'Accélérateur Linéaire*

# Strategy to measure the Higgs mass, width and invisible decays at ILC

François Richard, Philip Bambade

LAL, Univ Paris-Sud, IN2P3/CNRS, Orsay, France

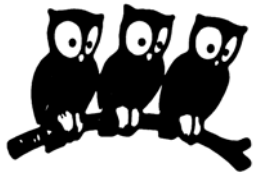

# Strategy to measure the Higgs mass, width and invisible decays at ILC

François Richard, Philip Bambade

LAL, Univ Paris-Sud, IN2P3/CNRS, Orsay, France

## Abstract

This document is meant to provide semi-quantitative arguments to evaluate the luminosity needed at ILC to achieve a precise measurement of the Higgs mass, width and invisible decays. It is shown that for $m_H$=120 GeV, one can save an order of magnitude on the luminosity needed to achieve a given precision on the Higgs mass, as compared to what can be obtained at √s=350 GeV, by running near threshold. Since the recoil mass resolution near threshold is independent of the Higgs mass, one can also access the Higgs width for masses above 170 GeV. This strategy of running just above threshold is also optimal to measure or set upper limits on the Higgs invisible branching ratio. Two MSSM scenarios are presented to illustrate the potential interest of an optimized recoil mass resolution. A simplified description of the various experimental mechanisms affecting this type of measurement is presented: detector resolution for leptons and jets, luminosity and beamstrahlung energy dependence, initial and final radiation of the involved leptons.

## 1. Introduction

ILC can provide a large sample of clean HZ events. It is therefore expected that with this sample and using the recoil mass to the Z, with Z decaying into $\mu^+\mu^-$ or $e^+e^-$, one can precisely reconstruct the Higgs boson mass, irrespective of its decay modes.

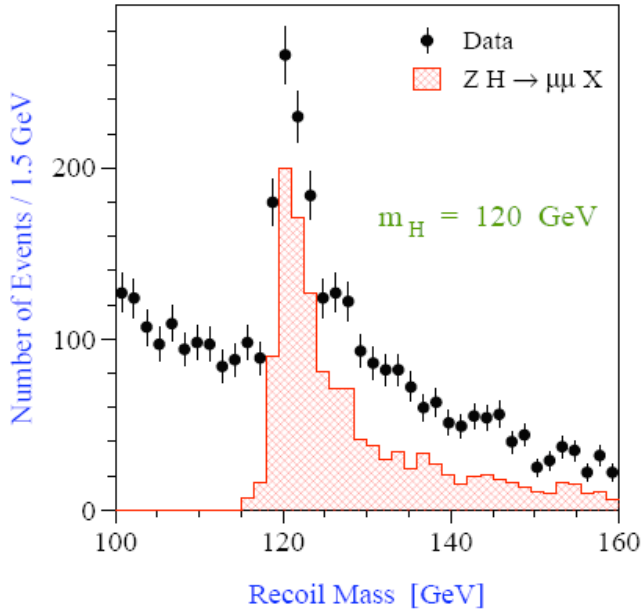

Figure 1: Recoil mass to Z boson for √s = 350 GeV [1].

This is illustrated in Figure 1, taken from TESLA TDR [1], where one assumes that the Z decays into muons.

In this note we would like to underline that to achieve the best mass resolution, one needs to operate close to the HZ threshold, contrary to the assumption of previous studies, like the one shown in Figure 1, which suggested using data taken at the top threshold.

Also, improved momentum resolution compared to that assumed in the TESLA TDR can be expected when using the TPC in conjunction with the external detectors (see Figure 6 in the Appendix).

## 2. Optimal working energy to measure the Higgs mass

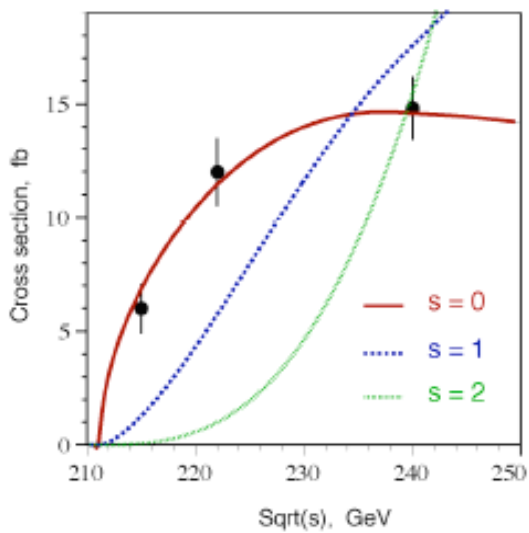

Figure 2: Threshold behaviour of Higgs boson production in association with a Z boson, for different assumptions on its spin [1].

We note that the mass resolution of about 1.5 GeV in Figure 1 is obtained at √s=350 GeV, which is well above threshold for the considered Higgs mass of 120 GeV. This means that the Z particle is boosted and therefore one finds that the error on the momentum of the muons p, which goes like $p^2$, is four times larger on average than near threshold (see Appendix).

Working near threshold would of course be unattractive if the rate were reduced. However this is not so since HZ is produced in an s-wave and therefore has a steep threshold as shown in Figure 2. Quantitatively, for a 120 GeV Higgs particle, one finds in fact that the production cross section is increased by about a factor two at 220-230 GeV with respect to its value at 350 GeV [1].

Finally one should note (see Appendix) that similar gains in resolution are obtained using Z hadronic decays, which further justifies operating near threshold.

## 3. Effect of beamstrahlung

One may object to above arguments that, in spite of an improved mass resolution, the net effect will be significantly attenuated by beamstrahlung (BS). The qualitative reply to this argument can be understood by noticing that, with the TESLA parameters at √s=500 GeV, on average more than 50% of the beam particles do not radiate at all before interacting while the rest radiate with strong peaking at small energies. One can therefore expect a mass distribution with a narrow peak reflecting the momentum resolution plus a BS tail carrying about 50% of the cross-section.

To see this, one can use for BS the parameterisation of the beam smearing given in reference [2]:

$$f(x) = a_0 \delta(1-x) + a_1 x^{a_2}(1-x)^{a_3}$$

with, for √s=500 GeV, $a_0$=0.53, $a_2$=13.895 and $a_3$=-0.63 while $a_1$ is adjusted to get a normalized distribution.

We have instead used the following formula provided by P. Chen and K. Yokoya [3]:

$$g(y) \sim \left[\frac{N_\gamma \kappa^{1/3} e^{-\kappa y}}{\Gamma(1/3) y^{2/3}} + \delta(y)\right] e^{-N_\gamma}$$

which gives the BS photon distribution when y=k/E<<1. For ILC (TESLA) at 500 GeV, the total number of emitted photons is $N_\gamma$=1.26(1.45) while $\Upsilon$=0.054(0.046) and $\kappa$=2/(3$\Upsilon$)=14.5(12.3). Note that the above formula would give $a_0$ well below 50%. To get the correct answer one should use the average number of radiated photons emitted before an interaction, which is about one half the total number used above.

The luminosity spectrum was obtained through the formulae in [3] assuming ILC parameters. Good agreement with the generator Guinea Pig [4] was found, as shown in Figure 7 in the Appendix. Using this spectrum and the mass resolution obtained in the Appendix, the differential cross-section for Hμμ was computed and is shown in Figure 3, at √s=230 GeV (blue curve) and 350 GeV (red curve), strikingly illustrating the improvement expected by running near threshold.

Table 1 below shows how this improvement translates into a gain in the luminosity needed to reach a given accuracy, 30 MeV, on the Higgs mass measurement.

| $E_{CM}$ | σ(Hμμ) fb No ISR | Average lepton momentum, GeV | $\sigma_{Mh}$ MeV | Muon Radcor | Electron Radcor | $\mathcal{L}$(30 MeV) fb$^{-1}$ μμ+ee |
|---|---|---|---|---|---|---|
| 350 | 4.6 | 83 | 900 | 0.45 | 0.27 | 780 |
| 230 | 9.1 | 54 | 200 | 0.36 | 0.18 | 20 |

Table 1 : Comparison of integrated luminosities at √s = 230 and 350 GeV needed to reach a 30 MeV precision on the Higgs boson mass. Also shown are the production cross sections and the average single event mass resolutions $\sigma_{Mh}$ and lepton momenta. Both internal and external radiative effects are taken into account in the evaluation (see Appendix).

*Additional considerations*

1/ The energy widths of the beams, which are at the 6 10$^{-4}$ level on average, would produce negligible effects given the single event recoil mass resolutions discussed. The mean energy would on the other hand need to be stable from bunch to bunch (or at least monitored) at the level of 1-2 10$^{-4}$ for 350-230 GeV, to avoid degrading the final mass resolution.

2/ Initial (ISR) and final state (FSR) radiation effects must be taken into account in the evaluation. The net effect is to reduce the peak by about $(2.35\sigma_{Mh}/E_{CM})^b$ where b is the equivalent radiator. The value of b is about 11.5% for ISR. For FSR, the value of b is about 6% for muons and 10.7% for electrons. Radiation in the detector can be estimated at the 4% level and also result in a lower efficiency (see Appendix). One could try to recover partially the loss from FSR by measuring the radiated photons in the ECAL but this can only work for photons below 500 MeV given the poor resolution of ECAL as compared to the tracking.

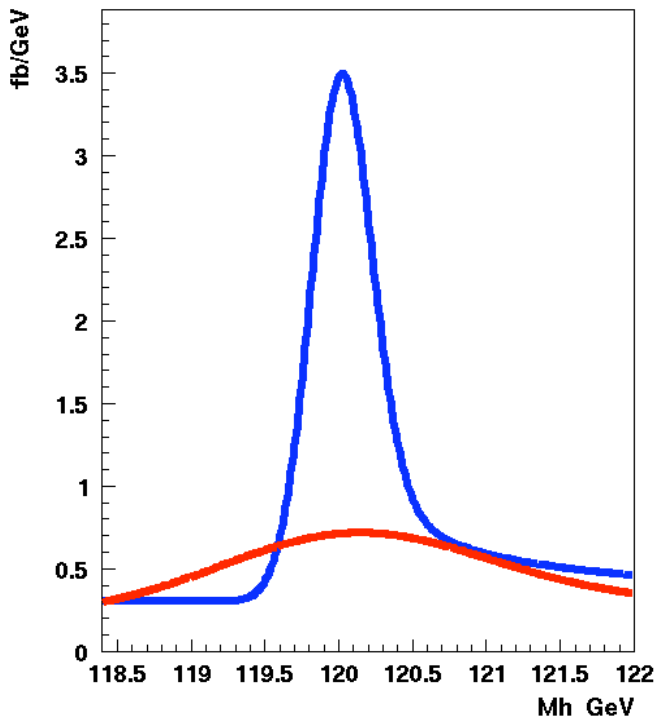

Figure 3: Differential cross-section for Hµµ at √s=230 GeV (blue curve) and 350 GeV (red curve).

3/ A 5-constraint kinematic fit, imposing energy-momentum conservation and the Z boson mass, would not help for leptonic decays, given that the resolution on the Z mass ~ $1/\sqrt{2}\sigma_p/p$ which for p~45 GeV and $\sigma_p/p^2 = 5.10^{-5}$ GeV$^{-1}$, gives 0.16% or 150 MeV, is well below the Z natural width. This remains true, given the assumed momentum resolution, even if one operates at √s=500 GeV. A 4-constraint kinematic fit using only energy-momentum conservation could on the other hand further improve the mass resolution but would of course depend on the Higgs boson decay mode.

## 4. Energy dependence of the Luminosity

<u>Beam parameter optimisation</u>: The evaluation in Table 1 implicitly assumes that the luminosity $\mathcal{L}$ does not depend on energy, which is not necessarily exactly the case. Generally speaking [5], optimising the luminosity of a linear collider in the presence of beam-beam effects leads to:

$$\mathcal{L} \sim \eta \frac{P_{Electrical}}{E_{CM}} \sqrt{\frac{\delta}{\varepsilon_{n,y}}} H_D$$

The effective beam power ηP (or at least its upper limit) and the normalized vertical emittance $\varepsilon_{n,y}$ can be considered fixed parameters for most linear collider designs. In our calculation we have also assumed that the beamstrahlung parameter δ could be maintained constant. This would imply a luminosity which increases at lower energy. In practice this is however not realistic since the number of particles per bunch N usually has an upper limit due to wake-field effects, or is limited within the damping rings. The beam power P is therefore proportional to energy and the overall luminosity constant as long as δ can indeed be maintained constant.

But is it possible to maintain a fixed δ ? This parameter goes like:

$$\delta = E_{CM} \frac{N^2}{\sigma_z \sigma_x^2} \quad \text{with} \quad \sigma_x^2 \sim \varepsilon_x \beta_x / \gamma$$

meaning that for fixed N and $\varepsilon_x$, the product $\beta_x \sigma_z$ should be reduced as the square of $E_{CM}$. If $\sigma_z$ is reduced, then one can also consider somewhat increasing the luminosity by reducing $\beta_y$ by a similar factor. This is so because the above mentioned luminosity optimisation normally assumes $\sigma_z \leq \beta_y$ to avoid optical depth of field effects in the vertical plane at the interaction point (the so-called hour-glass effect).

Using the Guinea Pig simulation we have checked that it is indeed possible to recover the luminosity reachable with nominal parameter at 350 GeV, either by decreasing $\beta_x$ by about a factor 2 with respect to the nominal value or, alternatively, by simultaneously decreasing $\beta_x$ and $\sigma_z$ by a factor 2/3 and $\beta_y$ by a factor 2.

With such parameters, beamstrahlung emission is maintained constant between 350 and 230 GeV. With the standard parameters, beamstrahlung emission would decrease, however the fraction of the luminosity within an energy bin corresponding to the narrow single-event Higgs boson recoil mass resolution would increase only by about 30% while the total luminosity would decrease by a factor 2. Hence increasing the luminosity by increasing the beamstrahlung is unexpectedly beneficial in this particular instance. Since the gain is moderate, such an optimisation is not absolutely crucial to the running scenario described here, which is also a good sign for the robustness of the results obtained.

While the proposed optimised parameters ($\beta_x$ ~10–15 mm, $\sigma_z$ ~ 0.2–0.3 mm and $\beta_y$ ~ 0.2–0.4 mm) are within the range considered for the ILC design, they are close to its edge. Practical feasibility should be explored further. Overall, there seems nonetheless to be enough flexibility to keep similar luminosity at 230 as at 350 GeV, and hence benefit from the larger cross section near threshold.

<u>Undulator-based positron production</u>: The operational scenario suggested in this paper also assumes that one can run at 230 GeV without limitations. This may not be exactly true in the undulator scheme, which uses electrons with at least 150 GeV to generate the positron beam. To be able to run near the Higgs boson production threshold as advocated, the electron beam must be decelerated from 150 GeV to about 115 GeV after its passage through the undulator. This may have practical implications (or even unwanted consequences) and is therefore important to analyse carefully. A solution should be found to enable running with reasonable performance near this limit, or else this could constitute an objection to the undulator idea itself, given the importance of achieving optimal running conditions for the HZ channel and the need for a threshold scan.

## 5. Measuring the SM Higgs boson width?

Much more challenging would be to measure the effect of the Higgs width on the Gaussian mass distribution. Two obvious difficulties are to be considered:

1/ The size of the width itself which, in the SM, is negligible for Higgs boson masses below about 170 GeV,

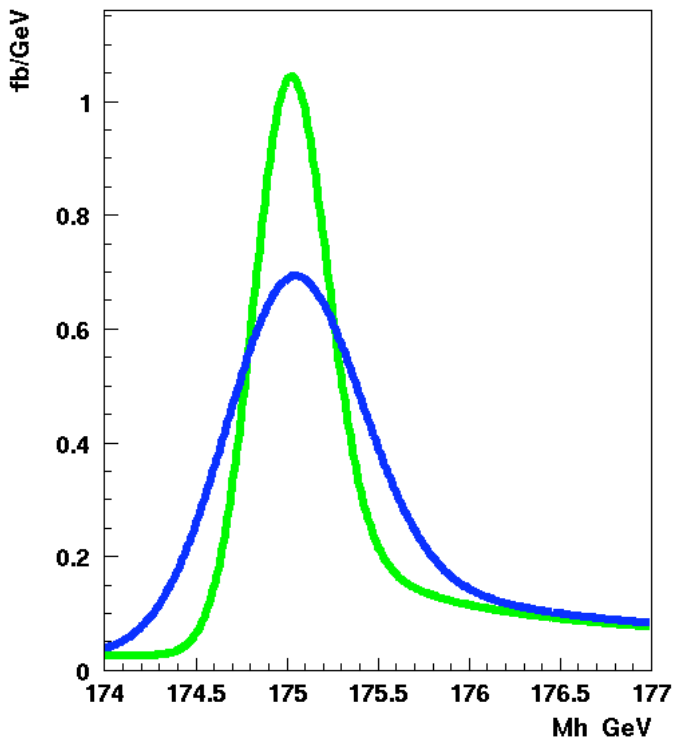

Figure 4 : Differential cross-section for Hμμ at √s=290 GeV for $M_H$=175 GeV with (blue) and without (green) including the Breit-Wigner width $\Gamma_H$=500 MeV.

2/ The effect of this width on the effective resolution, in case the two should be combined quadratically.

Empirically, one finds however that combining Gaussian and Breit-Wigner distributions with σ° and Γ widths, respectively, results in the following linear dependence for the total width σ of the combined distribution:

$$\sigma = \sigma° + 0.65*\Gamma/2$$

The possibility to measure the Higgs width is illustrated in Figure 4 for $\Gamma_H$=500 MeV, corresponding to $M_H$=175 GeV in the Standard Model. The picture compares the distributions obtained with (blue curve) or without (green curve) the Higgs width contribution.

Table 2 gives the integrated luminosity needed to measure the Higgs boson width to a 10% accuracy, assuming that $\sigma^0$ can be predicted accurately to deduce it from the overall width σ.

| ECM GeV | σ(Hμμ) fb | Average muon momentum,GeV | $\sigma_{Mh}$ MeV | $\mathcal{L}$ (10%) fb$^{-1}$ |
|---|---|---|---|---|
| 350 | 3.1 | 71.5 | 455 | 500 |
| 290 | 3.2 | 53 | 200 | 100 |

Table 2: Comparison of integrated luminosities at √s = 290 and 350 GeV needed to reach a 10% accuracy on the SM Higgs boson width, for $M_H$=175 GeV. Also shown are the production cross sections and the average single event mass resolutions $\sigma_{Mh}$ and lepton momenta.

It is remarkable that the recoil mass resolution at threshold does not increase with Higgs boson mass. This means that if such a heavy Higgs is produced at ILC, there is some hope to measure its width.

It should be noted that for such a measurement to be feasible, both the momentum resolution of the final state leptons and the differential luminosity spectrum must be known accurately. Running at the Z pole, where muon pairs are mono-energetic, allows a direct verification of the momentum resolution and also an absolute calibration of the momentum in the appropriate range. Concerning the luminosity spectrum, it can be precisely monitored using the acollinearity distribution [7] of Bhabha events.

## 6. An MSSM scenario

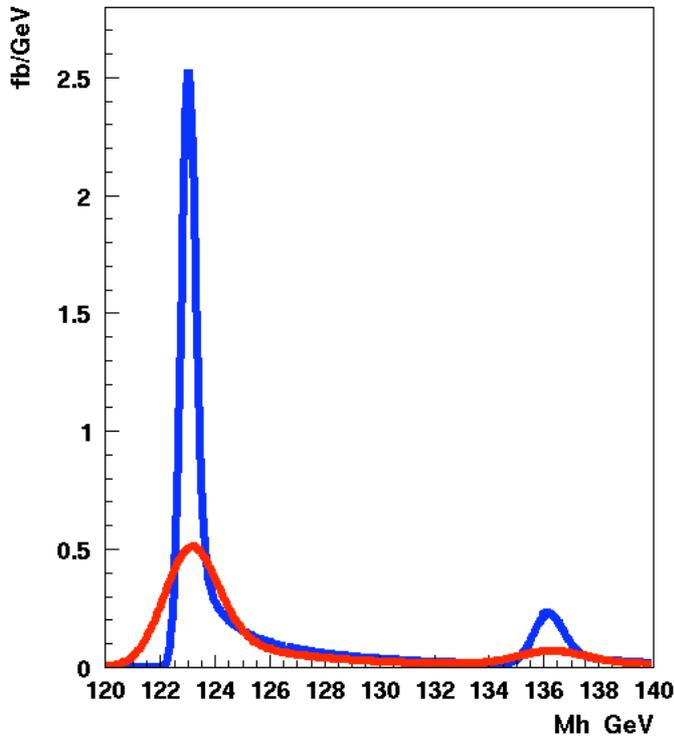

Figure 5: Differential cross-section for hµµ and Hµµ in the MSSM model described in [6], at √s=230 GeV (blue) and √s=350 GeV (red).

Within the Minimal Super Symmetrical Model (MSSM) the lightest Higgs boson has a mass of about 120 GeV and, in what is called the decoupling scenario, where the pseudo-scalar Higgs A is very heavy, behaves like the Standard Model Higgs, which means that its width is not measurable. There is however another possible scenario in which the two CP-even Higgs states, h and H, are almost degenerate in mass with A and can be observed simultaneously in the hZ and HZ channels. Moreover the couplings to b quarks can be at variance with the Standard Model prediction when tanβ becomes very large, thereby increasing the total width to a measurable value. Reference [6] proposes such a scenario, the main parameters of which are summarised in Table 3.

| Type | $\sigma_{hZ}$/SM | $\Gamma_{tot}$ GeV | Mass GeV |
|------|------------------|--------------------|----------|
| h    | 0.873            | 0.18               | 123      |
| H    | 0.127            | 0.873              | 136      |

Table 3: Main parameters of the MSSM scenario described in [6].

Figure 5 shows the expected behaviour of the reconstruction of the h and H Higgs boson masses in this model, at $E_{cm}$=230 GeV (blue curve) and at $E_{cm}$=350 GeV (red curve), illustrating both the potential of ILC in this scenario and the improvement expected at 230 GeV. Note that we assume that the ZZγ background can be removed by measuring the b-quark pair decays of the two Higgs particles, which represent about 90% of the branching ratios.

In this scenario, ILC can:

1/ Separate the two states and therefore measure the masses $M_h$ and $M_H$ and the couplings, which allows $\sin^2(\alpha-\beta)$ to be determined,

2/ Measure the widths, which provide the two quantities $\sin^2\alpha/\cos^2\beta$ and $\cos^2\alpha/\cos^2\beta$.

These six observables, where α and β are mixing angles, over-constrain the MSSM and allow a complete verification of this model. Deviations from MSSM predictions would establish some mixing effects with a non-minimal component, usually an isoscalar Higgs boson, or eventually other types of scalar like radions, as predicted within theories with extra dimensions.

## 7. A CP violation scenario

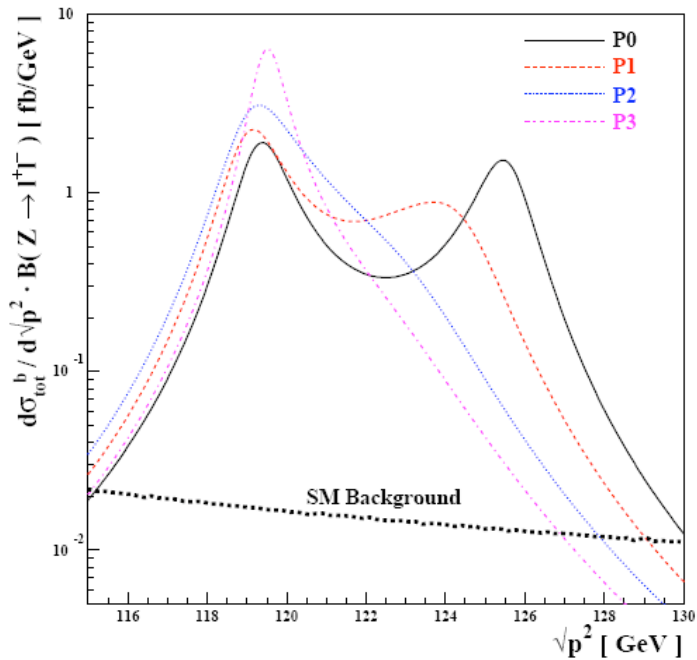

Figure 6: Differential cross-section for $hl^+l^-$ in the MSSM model with CP violation described in [7] for various mixing scenarios.

In the MSSM scenario there are various sources of CP violation (complex squark masses, gaugino masses and trilinear couplings) such that the three neutral Higgs bosons mix to mass eigenstates, $H_{1,2,3}$, of indefinite CP (instead of the usual h,H,A states). In [7] this case has been studied for an e+e- collider with the conclusion that one can use the high precision on the recoil mass to disentangle a scenario, expected for high values of tanβ, for which all three neutral Higgs states may have similar masses and strongly mix with each other.

This behaviour is shown in figure 6 taken from [7] which displays the complicated patterns occurring in various models. Improving by a factor 5 on the mass resolution, 1 GeV, used in [7] would certainly help greatly in separating the components.

# 8. Measuring the Higgs invisible decays

In several extensions of the Standard Model one predicts stable neutral light objects. The well known example is the case of the lightest neutralino which is expected to provide a good candidate for dark matter within supersymmetry. There are however many other examples of such particles, for instance in theories with extra dimensions or in other alternate scenarios like the Little Higgs scheme. One can also mention the so-called Goldstone particles, which appear in the context of symmetry violation, the best example being the lepton number violation related to neutrino masses (e.g. the Majorons).

Since a light Higgs boson mass, say below 150 GeV, couples very weakly to standard fermions (hence the very narrow width expected) there is a definite possibility that these new particles could significantly contribute to, or eventually entirely dominate, the Higgs particle decays. It is therefore essential to provide the highest sensitivity on the measurement on the Higgs invisible branching ratio $BR_{inv}$, as it carries a large discovery potential.

LHC can reach a 5% (13%) limit on $BR_{inv}$ with 100 fb$^{-1}$ (10 fb$^{-1}$) [9]. This may or may not be sufficient to see the manifestation of such models.

To understand the role of mass resolution, signal and background contributions one can write the significance of a signal as:

$$\frac{N_{Sig}}{\sqrt{N_{Bg}}} = \sqrt{\frac{L}{n\sigma_{Mh}}} \frac{BR_{inv}\varepsilon_{HZ}\sigma_{HZ}}{\sqrt{\varepsilon_{ZZ}\sigma_{ZZ}}}$$

This formula says that the luminosity L needed increases proportionally to the mass resolution and decreases quadratically with the HZ cross-section. The latter means that the hadronic Z decays provide the highest sensitivity. Taking a mass interval $\pm 2\sigma_{Mh}$ corresponds to n=4 in our formula. The results from [10] can then be easily reproduced with this expression with $\sigma_{Mh}$=7.3 GeV and $\varepsilon_{HZ}$=34%, which seems reasonable taking into account ISR and beamstrahlung effects.

The advantage of operating at 230 GeV rather than 350 GeV in the centre-of-mass is illustrated in Table 4. The semi-analytical evaluation of the single event recoil mass resolution presented in the Appendix was used. The contribution from the WW final states was neglected (it has been confirmed to be small [10]). Confusion with the standard HZ process when the Z boson decays into neutrinos is assumed to be negligible for a 120 GeV mass as long as an optimal energy flow can be used.

| $E_{cm}$ GeV | $\sigma(HZ_{had})$ fb (34% eff) | $\sigma(Z_{had}Z_{inv}+\gamma)$ fb $\pm 2\sigma_{Mh}$ | $\sigma_{Mh}$ GeV Hadrons | $\mathcal{L}$ fb$^{-1}$ 95% CL $BR_{inv}$<2% | $\mathcal{L}$ fb$^{-1}$ measure $BR_{inv}$=2±0.5% |
|---|---|---|---|---|---|
| 350 | 30 | 10 | 7.3 (1C fit) | 85 | 500 |
| 230 | 60 | 4 | 2.3 (1C fit) | 8 | 50 |

Table 4 : Integrated luminosities at √s = 230 and 350 GeV needed to set a 2% upper limit on $BR_{inv}$ at 95% CL and to measure a 2% value with a 0.5% precision, for $M_H$=120 GeV. Also shown are the effective production cross sections in the hadronic mode for a 34% experimental acceptance and the average single event mass resolutions $\sigma_{Mh}$.

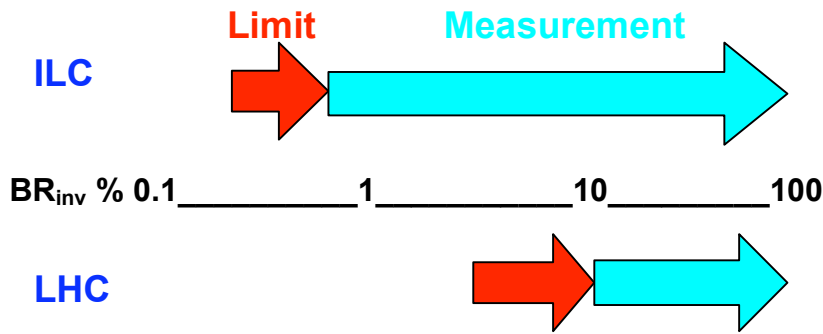

Above diagram compares the ILC and LHC performances for measuring or setting upper limits on $BR_{inv}$. It should also be stressed that:

1/ The mass of an invisibly decaying Higgs (and eventually its width) can also be precisely measured at ILC, which would not be the case at LHC,

2/ A branching ratio into fermions remains measurable at ILC down to a fraction of per cent. In this case, the Higgs boson width increases by two orders of magnitude and becomes also measurable even for $m_H$=120 GeV.

If the Higgs boson decays invisibly, but not dominantly ($BR_{inv}$ <99%), then one can estimate the invisible width due to new physics by comparing it to Standard Model visible decay branching ratio $BR_{vis}$. If the Higgs decays 100% invisibly, measuring the total width, which would be possible at ILC even for a 120 GeV mass, provides the only observable giving access to an estimate of the Higgs coupling to the invisible particles, an essential piece of information to understand the mechanism.

There are various experimental limitations, especially for very low $BR_{inv}$ values, which come into this analysis but we think that they can be handled using the data themselves. They can of course be studied in Monte Carlo analyses. Examples are: the recoil mass resolution, the background level. Both of these can be monitored using the large sample of ZZ data itself. The same should be true for minor backgrounds like WW and HZ itself.

In conclusion, the physics case for measuring or setting a limit on invisible decays is very high at the ILC, especially as this channel is much more difficult at the LHC. The range which can be covered is the widest if one can run near the HZ threshold.

## 9. Measuring Higgs hadronic modes

Until now we have worried about measuring the Higgs boson mass, width, invisible decay modes and we have found that for $M_H \sim$ 120 GeV, running near threshold rather than at √s=350 GeV saves about an order of magnitude on the luminosity needed. What about the other observables, in particular the various decay branching ratios?

The largest sample of ZH events comes from Z hadronic decays. When the Higgs boson decays hadronically into b- or c-pairs or into gluon pairs, one can use a kinematic fit with 5 constraints to achieve a good mass resolution, e.g. about 2(2.5) GeV at 230(350) GeV.

Before performing the fit, one can also reconstruct the hadronic final state to remove most of the ZZ$\gamma$ background. These events can fake a recoil mass consistent with the Higgs mass, while most of them have a measured hadronic mass peaking at the Z mass and can thus be easily eliminated given the hadronic mass resolution below 4 GeV. For this selection there is hence some advantage from the point of view of the reconstruction in operating at 350 GeV, since the direct mass resolution (meaning without use of a 5C fit) is improved by 20% allowing a slightly better background rejection.

Nonetheless, while a detailed Monte Carlo study would be needed for a more quantitative conclusion, the first guess is that the branching ratio measurements would still require, for a given precision, about twice less luminosity at 230 GeV due to the increase in HZ cross section and no very significant deterioration in mass resolution.

## 10. Summary, conclusions and prospects

In this note, using a simplified semi-analytical treatment, we have shown that the Higgs boson mass is best measured near the HZ threshold and that the recoil mass resolution does not depend on the Higgs mass. If there is a Higgs boson with a 120 GeV mass, one obtains the best use of luminosity by running at about $\sqrt{s}$= 230 GeV and by having available the best possible momentum resolution. This feature is also true for measuring or setting a limit on the invisible branching ratio where one needs good jet energy reconstruction. If invisible decays dominate the Higgs boson decays, ILC provides the unique feature of being able to measure the Higgs width and therefore the couplings to the invisible particles.

Running for an integrated luminosity 500 fb$^{-1}$ at $\sqrt{s}$=350 GeV, as usually considered [1], cannot be justified by the top quark threshold scan alone, as it requires much less luminosity nor, at present, by any predictable scenario.

In most cases an exploration for new physics would requires the highest luminosity being taken at $\sqrt{s}$=500 GeV. This is also true for important Standard Model measurements such as ttH and ZHH.

It appears however that such an energy does not provide appropriate conditions for studying ZH. Therefore, if LHC confirms the existence of a light Higgs, it will be essential to provide the possibility to run ILC at energies below $\sqrt{s}$=300 GeV with an optimal luminosity.

In the future one needs to perform a complete simulation of the Higgs channels at 230 GeV using the complete tools becoming available. It will also be important to investigate further the energy dependence of the luminosity (see Section 4), in particular to assess the feasibility of the suggested optimised beam parameters and of running below the energy required to produce positrons in the context of the undulator scheme.

## Acknowledgements

Useful discussions with K. Moenig are gratefully acknowledged. We also wish to thank Y. Mambrini and M. Berggren for providing, respectively, the detailed parameters of the MSSM study and an updated momentum resolution curve from the LDC concept study
We acknowledge the support of the European Community-Research Infrastructure Activity under the FP6 "Structuring the European Research Area" programme (CARE, contract number RII3-CT-2003-506395)

## Appendix

Mass resolution

1/ For leptonic decays of the Z boson, and taking that $\sigma_p/p^2=k$ ($k=5 \cdot 10^{-5}\,\text{GeV}^{-1}$ for the TESLA design), one easily derives the recoil mass error to the Z in the HZ mode:

$$\delta M_h^2 = \sqrt{2}kp(4Ep - M_Z^2)$$

where, near threshold, $2E \sim M_H + M_Z$ and the lepton momentum $p \sim M_Z/2$. On the other hand, rather quickly above threshold, $p \sim E/2$ where E is the beam energy and there is a fast deterioration of the mass resolution.

Instead of taking k constant one can use results recently obtained for the LDC [11] detector concept, which take into account the improved accuracies expected from the TPC and the impact of external silicon detectors. The angular dependence of this momentum resolution is not important and therefore has been neglected here.

One has:

$$p = \frac{4E^2 - M_h^2 + M_Z^2}{8E}$$

which for 2E=230(350) GeV gives a lepton momentum of 51(83) GeV for symmetric decays. From the above expression of the mass resolution, one therefore expects a substantial deterioration by going to 350 GeV. For asymmetric decays, this deterioration gets even worse, as it becomes dominated by the most energetic lepton:

$$\delta M_h^2 = kp(4Ep - M_Z^2)$$

The ratio $p_{max}/p$ varies between 1 and $1+\beta_Z \cos\theta$, where $\beta_Z$ =0.48(0.84) at 230(350) GeV. Keeping 70% of the events one finds that the mass resolution is increased by up to 80% at 350 GeV. In the results in Tables 1,2 and 4 we have given averaged resolutions with no restriction at 230 GeV and keeping only 70% of the acceptance at 350 GeV. These averaged resolutions have significant dispersions: 200±30 MeV at 230 GeV and 900±260 MeV at 350 GeV.

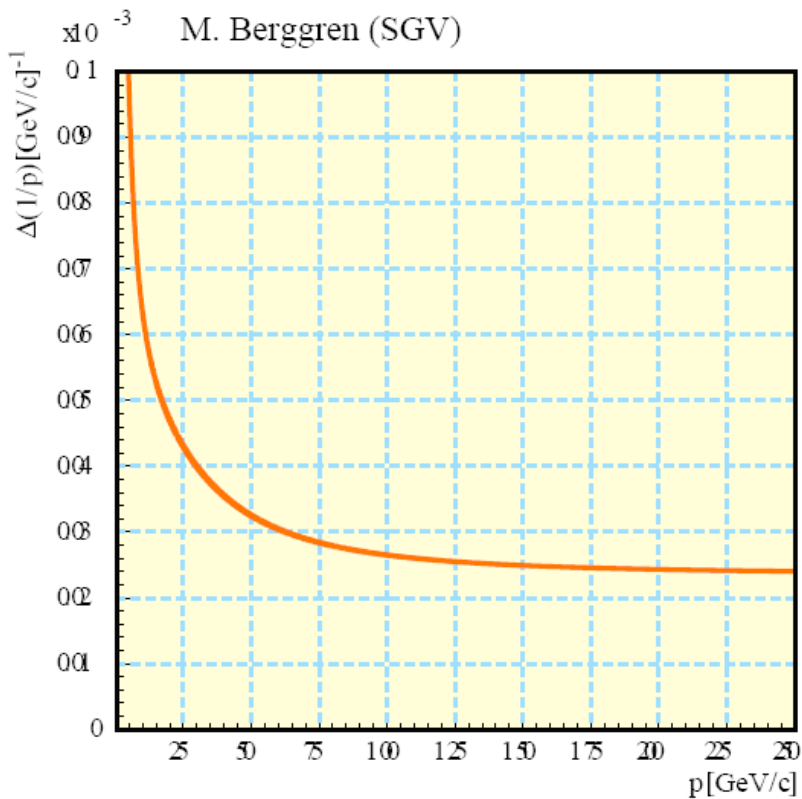

Figure 7: Momentum resolution for the LDC [10] detector concept with improved accuracies from the TPC and external silicon detectors.

2/ For hadronic decays of the Z boson one can use a kinematic fit with one constraint, imposing that:

$$M_Z^2 = 2pp'(1 - \cos\vartheta)$$

With this constraint and neglecting the angular errors and the Z width one has:

$$\delta M_h^2 = 4EA\sqrt{p}(1 - \frac{p'}{p})$$

where the calorimeter accuracy is dp=A√p with A ~ 0.3. Again, the largest errors come from asymmetric decays. To minimize this error one should determine p, the highest momentum, from calorimetry and deduce p' from the 1C constrain. Near threshold, p ~ p' and therefore the above error becomes very small.

There is an additional error due to the Z natural width:

$$\delta M_h^2 = \frac{\delta M_Z^2}{M_Z^2}(M_Z^2 - 4Ep)$$

Recall that this effect has to be linearly added to the preceding term with the empirical formula given in the main text:

$$\sigma = \sigma° + 0.65*\Gamma/2$$

At 230 GeV, one finds an averaged mass resolution of 2.3±0.8GeV (the variation corresponds to the spread in momentum of the quarks). At 350 GeV, the same calculation gives a resolution of 7.3±2.3 GeV. Note that in the latter case one needs to measure jets energies up to 150 GeV which may be challenging for the particle flow method [12]. We have nevertheless kept the same coefficient A=0.3 in the whole momentum range.

3/ For Z and Higgs hadronic final states, one can perform a 5C fit which gives ~2.5(2)GeV mass resolution on the Higgs mass, at 350(230)GeV centre of mass energy.

### Radiative effects

There are four different radiative effects to be considered: beamstrahlung, ISR, FSR and external radiation for final state electrons.

1/ For beamstrahlung, we have cross-checked the P. Chen parameterisation in [3] with Guinea-Pig simulation [4] performed at √s=220 GeV. The agreement is reasonably good (see Figure 7).

2/ ISR and FSR can be simply described through the probability distribution:

$$dP = b\frac{dx}{x^{1-b}}$$

where x=k/E is the ratio between the photon and the lepton energies and:

$$b = \frac{2\alpha}{\pi}\left[\log\left(\frac{M^2}{m^2}\right) - 1\right]$$

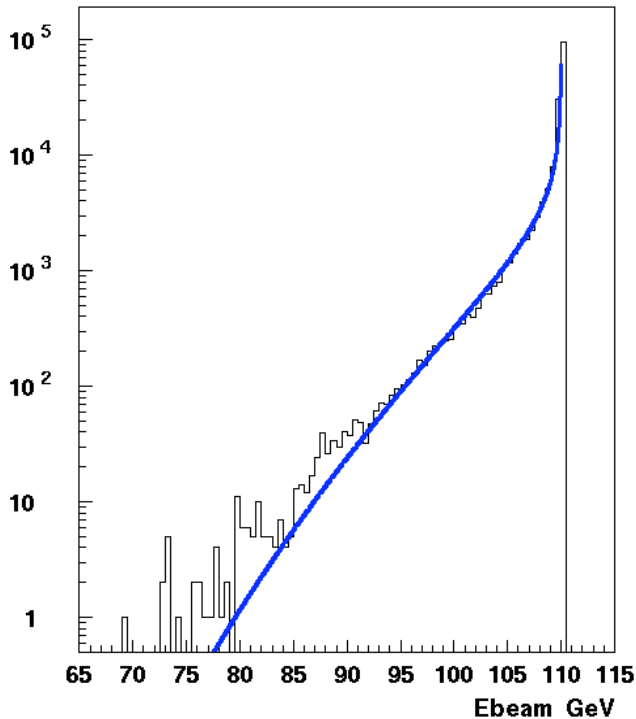

Figure 8: Differential spectrum for interacting beam particles computed with the Guinea-Pig simulation [4] and analytically using the expressions in [3].

where m is the lepton mass and M is the lepton pair mass. For ISR one has simply M²=s and m is the electron mass. For FSR M=$M_Z$. Remarkably enough one cannot neglect muon FSR, for which b=5.8%, with respect to electron FSR, for which b=10.8%.

3/ External radiation is only relevant for electrons and is governed by the detector properties. For the purpose of this study we have assumed that an electron will see, on average, d=3% of $X^0$ before entering the TPC. We have then added this contribution to the FSR for the electrons, taking b=2(4d/3$X^0$) = 8% to account for both leptons. This figure shows that the detector properties amplify significantly the losses due to FSR. The net result, shown in Table 1, is a reduction in efficiency of a factor of about two for electrons with respect to muons.

In practice one should convolute the Gaussian (or Breit Wigner) shape of the mass resolution with the probability distribution. This gives for the radiation effect the following attenuation on the peak:

$$P = \left(\frac{\Gamma}{M}\right)^b = \left(\frac{2.35\sigma}{M}\right)^b$$

where M=$E_{cm}$ for ISR and M=$M_Z$ for FSR. This formula can be intuitively derived by integrating the probability distribution given above between 0 and $\Gamma$/2.

<u>The ZZ$\gamma$ background</u>

The ZZ process dominates the background to the HZ channel if one uses the recoil mass method with lepton pairs. In the absence of radiation, only ZZ* can contaminate HZ with $M_{Z*}$=$M_H$. This contribution is however typically one order of magnitude below the contribution from ZZ$\gamma$, where the photon comes either from ISR or from beamstrahlung. This process constitutes a significant background when the energy of this photon is such that one of the Z boson combinations peaks at the Higgs boson mass.

The various components of the ZZ background, for the two energies discussed in the paper, are given in Table 5. For what concerns this background one sees that reducing further the beamstrahlung would not help much near threshold.

| Energy GeV | ZZ*(120) | ZZ$\gamma$ RC | ZZ$\gamma$ BS | Sum fb/GeV |
|---|---|---|---|---|
| 230 | 1 | 6.7 | 0.2 | 7.9 |
| 350 | 0.7 | 3.1 | 2.1 | 5.9 |

Table 5: Expected backgrounds for HZ(120 GeV), in fb/GeV, coming from the ZZ process, at the two energies. The ZZ*(120) background gives a virtual Z at 120 GeV. In the other two processes there is a photon (from initial radiation or beamstrahlung) emitted in the beam pipe with two Z bosons on mass-shell such that the recoil mass to the Z decaying leptonically peaks at 120 GeV.